\documentclass[letterpaper]{article}

\usepackage{natbib}  %

\usepackage{fullpage}
\usepackage[utf8]{inputenc} %
\usepackage[T1]{fontenc}    %
\usepackage{hyperref}       %
\usepackage{booktabs}       %
\usepackage{amsfonts}       %
\usepackage{nicefrac}       %
\usepackage{microtype}      %
\usepackage{amsmath,amsthm}
\usepackage{thmtools, thm-restate}
\usepackage{algorithm,algorithmicx,algpseudocode}
\usepackage{xspace}
\usepackage{color}
\usepackage{mathtools}
\usepackage{subcaption}
\usepackage{enumitem}
\usepackage{float}

\newcommand{\1}{\mathbb{I}}
\newcommand{\PP}{\mathbb{P}}
\newcommand{\E}{\mathbb{E}}

\title{Convergence Analysis of No-Regret Bidding Algorithms in Repeated Auctions}

\author{%
	Zhe Feng \thanks{Email: \texttt{zhe\_feng@g.harvard.edu}. Harvard University.} \and
	Guru Guruganesh \thanks{Email: \texttt{gurug@google.com}. Google.} \and 
	Christopher Liaw \thanks{Email: \texttt{cvliaw@cs.ubc.ca}. University of British Columbia.} \and 
	Aranyak Mehta \thanks{Email: \texttt{aranyak@google.com}. Google.} \and
	Abhishek Sethi \thanks{Email: \texttt{arsethi@google.com}. Google.}
}
\date{September 13, 2020}

\newcommand{\bE}{\mathbb{E}}

\newcommand{\eps}{\varepsilon}

\newtheorem{theorem}{Theorem}[section]
\newtheorem{lemma}[theorem]{Lemma}
\newtheorem{claim}[theorem]{Claim}

\newtheorem{assumption}[theorem]{Assumption}
\newtheorem{definition}[theorem]{Definition}

\newcommand{\AlgorithmName}[1]{\label{alg:#1}}

\newcommand{\SectionName}[1]{\label{sec:#1}}

\newcommand{\TheoremName}[1]{\label{thm:#1}}
\newcommand{\Theorem}[1]{Theorem~\ref{thm:#1}}

\newcommand{\AppendixName}[1]{\label{app:#1}}
\newcommand{\Appendix}[1]{\autoref{app:#1}}

\newcommand{\expects}[2]{\bE_{#1} \left[#2\right]}

\usepackage[colorinlistoftodos,textsize=tiny,textwidth=2cm,color=red!25!white,obeyFinal]{todonotes}

\newcommand{\comment}[1]{}
\newcommand*{\NOCOMMENT}{}%
\ifdefined\NOCOMMENT
    \newcommand{\ggnote}[1]{}
    \newcommand{\chris}[1]{}
    \newcommand{\amnote}[1]{}
    \newcommand{\zf}[1]{}
    
\else
    \newcommand{\ggnote}[1]{\todo[inline,color=blue!25!white]{GG: #1}}
    \newcommand{\chris}[1]{\todo[inline,color=green!20]{\textbf{Chris:} #1}}
    \newcommand{\zf}[1]{\todo[inline,color=cyan!25!white]{\textbf{ZF:} #1}}
    \newcommand{\amnote}[1]{\todo[inline,color=red!25!white]{\textbf{AM:} #1}}
\fi

\newcommand{\eat}[1]{}

\begin{document}
\maketitle

\begin{abstract}
The connection between games and no-regret algorithms has been widely studied in the literature. A fundamental result is that when all players play no-regret strategies, this produces a sequence of actions whose time-average is a coarse-correlated equilibrium of the game. However, much less is known about equilibrium selection in the case that multiple equilibria exist.

In this work, we study the convergence of no-regret bidding algorithms in auctions. Besides being of theoretical interest, bidding dynamics in auctions is an important question from a practical viewpoint as well. We study repeated game between bidders in which a single item is sold at each time step and the bidder's value is drawn from an unknown distribution. We show that if the bidders use any mean-based learning rule then the bidders converge with high probability to the truthful pure Nash Equilibrium in a second price auction, in VCG auction in the multi-slot setting and to the Bayesian Nash equilibrium in a first price auction. We note mean-based algorithms cover a wide variety of known no-regret algorithms such as Exp3, UCB, $\epsilon$-Greedy etc. Also, we analyze the convergence of the individual iterates produced by such learning algorithms, as opposed to the time-average of the sequence. Our experiments corroborate our theoretical findings and also find a similar convergence when we use other strategies such as Deep Q-Learning. 
\end{abstract}

\section{Introduction} \SectionName{intro} 

The connection between Learning and Games has proven to be a very innovative, practical, and elegant area of research (see, e.g.,~\cite{FL-book,CL-book,NRTV07}). Several fundamental connections have been made between Learning and Game Theory.  A folklore result is that if all players play low-regret strategies in a repeated general-sum game, then the time-averaged history of the plays converges to a coarse correlated equilibrium (see, e.g.,~\cite{blum-mansour}). Similarly, low-swap-regret play leads to correlated equilibria. 

In this work, we are interested in the setting of multi-agent learning in auction environments. Given the importance of auctions and bidding in the online advertising ecosystem,  this is an important question from a practical point of view as well. In this setting, an auction takes input bids and determines the allocation of ad-slots and the prices for each advertiser. This is a highly repeated auction setting over a huge number of queries which arrive over time. Advertisers get feedback on the allocation and cost achieved by their chosen bidding strategy, and can respond by changing their bids, and often do so in an automated manner (we will assume throughout that advertisers are {\it profit-maximizing}, although there are potentially other goals as well). As each advertiser responds to the feedback, it changes the cost landscape for every other competing advertiser via the auction rules. As advertisers respond to the auction and to each other, the dynamics lead them to an equilibrium. This results in the following fundamental question: Given a fixed auction rule, do such bidding dynamics settle in an equilibrium and if so, what equilibrium do they choose.

Surprisingly, neither the Auction Theory literature from Economics nor the literature on Learning in Games provides a definitive answer. Consider the simplest setting: a repeated auction for a single item, where the allocation is either first-price or second-price. Auction Theory suggests that bidders converge to {\em canonical equilibria} (see, e.g.,~\cite{krishna-book}): 
\begin{itemize}[leftmargin=*,itemsep=2pt]
\item  For a second-price auction (or more generally a VCG auction), bidders will choose to be truthful (bid their true value every time) as this strategy weakly dominates every other strategy, i.e., no other strategy can yield more profit. This is a weakly dominating strategy Nash equilibirum (NE).
\item For a first price auction in which each advertiser's value is picked from some commonly known distribution in an i.i.d.~manner, each advertiser will underbid in a specific way to achieve a {\em Bayesian Nash Equilibrium} (BNE). For example, when there are two bidders and the value distribution is the uniform distribution on $[0,1]$, then each advertiser will set its bid to be half of its true value.
\end{itemize}

While these canonical equilibria make intuitive sense, they are not the only equilibria in the respective games. For example, consider a single-item setting in which bidder 1 has a value of $1.0$ and bidder 2 has a value of $0.5$ for the item. While truthful bidding is an NE, any two values $b_1, b_2 \in [0.5, 1.0]^2$, with $b_1 > b_2$ also form an NE (in fact, an Envy-Free NE). Thus, there are an infinite number of NEs,  with very different revenues. Similarly, in the Bayesian setting where the values are drawn from say, a uniform distribution, there are many NEs as well. For example, one player could bid $1$ and the other player always bids $0$ regardless of their valuations. This issue of multiple equilibria is treated in the Economics literature via various notions of equilibrium selection but it is not clear if such selection occurs naturally in settings such as ours, especially via bidding dynamics.

To take the Learning in Games approach to answering the question, we have to fix the bidding dynamics. We assume bidders use no-regret (mean-based) online learning algorithms; these are natural and powerful strategies that we may expect advertisers to use. A lot of commonly used no-regret learning algorithms, e.g. multiplicative weights update (MWU), follow the perturbed leader (FTPL), EXP3, UCB, and $\eps$-Greedy, are all special cases of mean-based no-regret learning algorithms.

Indeed, there has been considerable work recently which studies various questions in the online advertising setting under this assumption (see, e.g., \cite{Tardos15}). Folklore results in Learning imply that under low-regret dynamics the time-average of the advertisers' bids will converge to a {\em coarse correlated equilibrium (CCE)} of the underlying game. However, there could be many CCEs in a game as well. Since every NE is a CCE as well, the above examples hold for the second-price auction. For the first price setting (even when the values are drawn uniformly at random) another CCE is for the two bidders to bid $(v_1+v_2)/2 + \epsilon$ and $(v_1+v_2)/2$ where $v_1$ and $v _2$ are the drawn values, and $v_1 > v_2$. Further, since any convex combination of these CCEs is a valid CCE (the set of CCE forms a polytope), there is an infinite number of CCEs in both first and second price auctions. So again we are left with the prediction dilemma: it is not clear which of these CCEs a low-regret algorithm will converge to. Some of the CCEs are clearly not good for the ecosystem, for revenue or efficiency. Further, as we elaborate below, even if there is a unique CCE in the games of interest, the convergence guarantee is only for the time-average rather than point-wise.

\subsection*{Questions} These examples directly motivate the questions we ask in this paper:
\begin{itemize}[leftmargin=*,itemsep=2pt,topsep=0pt]

\item  If the values are drawn in a bayesian setting, and the bidders follow a low-regret learning algorithm in a repeated second-price auction, do they converge to the truthful equilibrium?  

\item Similarly, in a first price auction with i.i.d.~values with bidders values drawn from a uniform distribution $[0,1]$, do they converge to the Bayesian Nash equilibrium (with two bidders) of bidding half of true value?

\item Do the bidding dynamics converge to such an equilibrium point-wise, i.e., in the last-iterate sense, or only in the time-average sense?  

\item When there are multiple slots, do the bidders converge to truthful equilibrium under VCG settings? 

\end{itemize}

Given the current state of the literature, we see these as fundamental questions to ask. The only guarantees we have are those known for general games:  Low-regret dynamics converge to some CCE and there is no guarantee for the last-iterate convergence. If it is the case that only the time average converges, then that means that bidders may be required to keep changing their bidding strategy at every time step (see the discussion on the non-point-wise-convergence results in~\cite{BP18} below), and would achieve very different rewards over time. This would not be a satisfactory situation in the practical setting.

\subsection*{Our Results} 

\zf{Cut the model, since we have the clear model in section 2.}

Our main result is that when each of the bidders use a mean-based learning rule (see Definition~\ref{def:mbl}) then all bidders converge to truthful bidding in a second-price auction and a multi-position VCG auction and to the Bayes Nash Equilibrium in a first-price auction.
\newtheorem*{itheorem}{Informal Main Theorem}
\begin{itheorem}
\TheoremName{meanBasedSPA_informal}
Suppose $n$ players whose value are drawn from a distribution over space $\{ \frac{1}{H}, \frac{2}{H}, \dots, 1 \}$ bid according to a mean-based learning algorithm in either (i) a second-price auction; (ii) a first-price auction (for $n = 2$ and uniform value distribution); or (iii) a multi-position VCG auction.
Then, after an initial exploration phase, each bidder bids the canonical Bayes Nash equilibrium with high probability.
\end{itheorem}
The formal statement of this theorem appears in \Theorem{convergence-spa}, \Theorem{convergence-fpa}, and \Theorem{convergence-multi-position} for second-price auctions, first-price auctions, and multi-position VCG auctions, respectively. Moreover, we show each bidder converges to bid canonical Nash Equilibrium point-wise for any time, which is in sharp contrast with previous time-average convergence analysis.

Throughout this paper, we assume the learning algorithms that the bidders use may be oblivious to the auction format that is used by the seller.

We complement these results by simulating the above model with experiments. In particular, 
we show that these algorithms converge and produce truthful (or the canonical) 
equilibria. Furthermore, we show that the algorithm converges much quicker than the theory would predict. These results indicate that these low-regret algorithms are quite robust in the context of Auctions.

\subsection{Our Techniques}
Our proof techniques involve a few key observations. For second-price and VCG auctions, we 
want to show that the bidding converges to truthful bidding. Firstly, note that if the other bidders bid completely randomly then the truthful arms have a slight advantage in profit. However, the other bidders themselves bid according to their own instance of the low-regret algorithms, thus the environment that a given bidder sees is not necessarily random; hence we need more insight. Fix a particular bidder, say bidder 1. In the beginning of the learning algorithm, all bidders do bid randomly. The next observation is that if the other bidders {\em happened} to converge to truthful bidding, then again bidder 1 will see completely random bids, because the other bidders' values are picked randomly at each stage. Hence we can say that both in the beginning and also if other bidders happen to converge to (or for some reason were restricted to) truthful bidding, then bidder 2 will see an advantage in truthful bidding and converge to that. It remains to show that in the interim period, when all bidders are learning, exploring, and exploiting, the truthful strategy builds and retains an advantage.

In a first price auction, the proofs follow the same structure.
However there are some technical difficulties that one must overcome.
Initially, both bidders bid uniformly at random and it is not difficult to show that bidding according to the canonical NE gives an advantage.
If the bidders happen to converge to the BNE then, of course, bidding according to the BNE is the optimal strategy for either player.
It is not clear, however, that when the opposing bid is \emph{not} uniform or bidding according to the BNE that an advantage is maintained.
Our technical contribution here is to show that an advantage for bidding the BNE is maintained which allows both bidders to converge to the BNE.

Our results show that we can achieve high probability results we can show that the model will bid truthfully \emph{for all time} (assuming a modest exploratory period in the begining). This requires a new partitioning argument which enables us to apply concentrations results for all times. This technique may be of independent interest. 

\subsection{Related Works} 
Our work lies in a wide area of the inter-disciplinary research between mechanism design and online learning algorithms, e.g.,~\cite{Auer,CL-book,blum-mansour}, and we only point out a  few lines of research which are more closely related to our work.

In online advertising, the setting where bidders may be running no-regret algorithms rather than best response, has recently been investigated and has garnered a significant amount of interest. For example, \citet{Tardos15} study how to infer advertisers' values under this assumption. However,~\citet{BP18} show, somewhat surprisingly, that even in very simple games (e.g., a zero-sum matching pennies game), the no-regret dynamics of MWU do \emph{not} converge, and in fact the individual iterates tend to {\it diverge} from the CCE. On the other hand, recent results in~\cite{daskalakis18ITCS,mertikopoulos} show that certain Optimistic variants of Gradient Descent and Mirror Descent converge in the last-iterate to the NE in certain zero-sum games. Our result can be seen as a contribution in this stream of work as well, in that we show that for the (non-zero sum) games arising from auctions that we study, 
mean-based learning algorithms converge in the last iterate, and, in fact, to the natural (Bayes) NE. On a related note, ~\citet{papa-pillouras} shows there is a conflict between the economic solution concepts and those predicted by Learning Dynamics. In that framework, one can consider this work as suggesting that perhaps there is no such conflict between economic theory and learning, in the context of games arising from auctions, as learning converges to the solutions predicted by auction theory.

Our work is also related with \emph{Learning to bid} literature,~e.g.,~\cite{Weed16,Feng18,Balseiro19}, where these papers focus on designing a good learning algorithm for the bidders in repeated auctions. In addition, ~\citet{Braverman18} considers how to design a mechanism to maximize revenue against bidders who adopt mean-based learning algorithms. In contrast, the auctions are fixed in our setting and we are interested in understanding the bidder dynamics.

Last but not least,~\citet{FeldmanLN} characterize multiple equilibria (NE, CE, and CCE) in first price auctions, under the prior-free (non-Bayesian) setting, and study the revenue and efficiency properties of these equilibria. They show there are auctions in which a CCE can have as low as $1-2/e \simeq 0.26$ factor of the second highest value (although not lower), and there are auctions in which a CCE can have as low as $0.81$ of the optimal efficiency (but not lower). However, our results show that even though there may be ``Bad'' CCEs, the natural dynamics do not reach them, and instead, converge to the classic canonical Nash equilibrium. 

\section{Model and Notations}
We consider the setting that there is a single seller repeatedly selling one good to $n$ bidders per round. At each time $t$, each bidder $i$'s valuation $v_{i,t}$ is i.i.d.~drawn from an unknown (CDF) distribution $F_i$ and bidder $i$ will submit a bid $b_{i,t}$ based on $v_{i, t}$ and historical information. In this paper, we assume the value and the bid of each bidder at any time are always in a $\frac{1}{H}$-evenly-discretized space $V = \{\frac{1}{H}, \frac{2}{H},\cdots, 1\}$, i.e, $v_{i, t}, b_{i, t}\in V, \forall i, t$. Let $v_ t = (v_{1, t}, \cdots, v_{n, t})$ be the valuation profile of $n$ bidders at time $t$, $v_{-i, t}$ be the valuation profile of bidders other than $i$, and similarly for $b_t$ and $b_{-i, t}$. Let $F$ be the (CDF) distribution of $v_t$ and $f_i$ be the probability density function (PDF) of bidder $i$'s value. Denote $m_{i, t} = \max_{j\neq i} b_{j, t}$ as the maximum bid of the bidders other than $i$ and $z_{i, t} = \max_{j\neq i} v_{j, t}$ be the maximum value of the bidders other than $i$. We denote $G_{i}$ as the (CDF) distribution of $z_{i, t}$ and $g_i$ as the associated PDF. For theoretical purpose, we propose an assumption about $G_i$ in the following,

\begin{assumption}[Thickness Assumption of $G_i$]\label{assumption:thickness-assumption}
There exists a constant $\tau > 0$ (may depend on $n$), s.t., $g_i(v) \geq \tau, \forall i\in [n], v\in V$. Without loss of generality\footnote{It is without loss of generality, since if $\tau > \frac{1}{H^{n-1}}$, we redefine $\tau :=\min\left\{\tau, \frac{1}{H^{n-1}}\right\}$.}, we assume $\tau \leq \frac{1}{H^{n-1}}$.
\end{assumption}

We assume the each bidder runs a no-regret learning algorithm to decide her bid at each time. Specifically, in this paper, we are interested in a broad class of no-regret learning algorithm known as mean-based (contextual) learning algorithm~\cite{Braverman18};
these include the multiplicative weights update (MWU), Exp3, and $\eps$-Greedy algorithms as special cases.

In this paper, we focus on the contextual version of mean-based learning algorithms, which can be used to model learning algorithms of the bidders in repeated auctions, defined in the following.
\begin{definition}[Mean-based Contextual Learning Algorithm]\label{def:mbl}
Let $r_{a, t}(c)$ be the reward of action $a$ at time $t$ when the context is  $c$ and $\sigma_{a,t}(c) = \sum_{s=1}^{t} r_{a,s}(c)$. An algorithm for the
contextual bandits problem is $\gamma_t$-mean-based if it is the case that whenever
$\sigma_{a, t}(c) < \sigma_{b, t}(c) - \gamma_t t$, then the probability $p_{a,t}(c)$ that the algorithm plays action $a$ on round $t+1$, given context $c$, is at most $\gamma_t$. We
say an algorithm is mean-based if it is $\gamma_t$-mean-based for some $\gamma_t$ such that $\gamma_t t$ is increasing and $\gamma_t \rightarrow 0$ as $t \rightarrow \infty$. \footnote{The mean-based learning algorithms proposed by~\cite{Braverman18} set $\gamma_t$ be a constant, which only depends on total number of time steps $T$. Here we extend it to be a time-dependent variable, which is used to show our \emph{anytime} convergence results.}
\end{definition}

In the repeated auctions setting, the context information received by each bidder $i$ at time $t$ is the realization of the valuation $v_{i, t}$. The reward function $r^i_{b, t}$ for bidder $i$ can be defined as
\begin{eqnarray}
\forall v \in [0, 1], r^i_{b, t}(v) := u_{i, t}((b, b_{-i, t}); v),
\end{eqnarray}
where $u_{i, t}((b, b_{-i, t}); v)$ is the utility of bidder $i$ at time $t$ when the bidder $i$ bids $b$ and the others bid $b_{-i, t}$, if bidder $i$ values the good $v$. 

\subsection{Learning Algorithms of Mean-Based Bidders}
In this paper, we focus on the setting where each bidder $i$ runs a $\gamma_t$-means-based contextual learning algorithm to submit the bid\footnote{Indeed, our analysis can be extended to the setting where each mean-based bidder has different $\gamma_t$ parameters. We assume they share parameters $\gamma_t$, for notation simplicity.}. In addition, we assume each bidder runs several pure exploration steps in the beginning to estimate the reward of each action (bid) for each context (value). We assume each bidder runs $T_0$ pure exploration steps: at each pure exploration step, each bidder $i$ uniformly generates a bid from $B$ at random, regardless of the realization of value $v_{i, t}$. To summarize, we describe the learning algorithm of mean-based bidders in Algorithm~\ref{alg:mbl}.

\begin{algorithm}
\caption{Mean-based (Contextual) Learning Algorithm of Bidder $i$}
\AlgorithmName{MBL}
\label{alg:mbl}
\begin{algorithmic}[1]
\Procedure{MBL}{$\gamma_t, T_0$}
\For{$t = 1, 2, \ldots, T_0$}
\State Choose bid $b_{i,t}$ uniformly from $V$ at random.
\EndFor
\For{$t=T_0 + 1, T_0 + 2, \ldots$}
\State Observes value $v_t$.
\State Choose bid $b_{i, t}$ following a $\gamma_t$-mean-based learning algorithm. 
\EndFor
\EndProcedure
\end{algorithmic}
\end{algorithm}

In the \emph{learning to bid} literature, there are different feedback models: full information feedback~\cite{papa-pillouras}, bandit feedback~\cite{Weed16,Feng18}, or cross-learning feedback~\cite{Balseiro19}.
However, our results hold for any feedback model, as long as each bidder uses the general mean-based learning algorithm to bid, shown in Algorithm~\ref{alg:mbl}.

\section{Second Price Auctions with Mean-based Bidders}

In this section, we analyze the learning dynamics of mean-based bidders in (repeated) second price auctions.
In second price auctions, the utility function of each bidder $i$ at time $t$ can be represented as,
\begin{eqnarray}\label{eq:utility-spa}
u_{i, t}((b, b_{-i, t}); v) = (v - m_{i, t})\cdot \1\{b \geq m_{i, t}\}
\end{eqnarray}

Since the bids of each bidder are in a discrete space, we break ties randomly throughout this paper.
We first show the following main theorem in this section, which proves that the mean-based learners converge to truthful reporting point-wisely, in the repeated second price auctions.

\begin{theorem}\label{thm:convergence-spa}
Suppose assumption~\ref{assumption:thickness-assumption} holds and $T_0$ is large enough, such that $\exp\left(-\frac{\tau^2 T_0}{32n^2H^2}\right) \leq \frac{1}{2}$ and $\gamma_t \leq \frac{\tau}{8nH}, \forall t \geq T_0$. Then at time $t > T_0$, each $\gamma_t$-mean-based learner $i$ will submit $b_t = v_{i, t}$ in repeated second price auctions with probability at least %
$p(t) = 1 - H\gamma_t - 4\exp\left(-\frac{\tau^2 T_0}{32n^2H^2}\right)$, for any fixed $v_{i, t}$. Note $p(t) \rightarrow 1 - 4\exp\left(-\frac{\tau^2 T_0}{32n^2H^2}\right)$ when $t \rightarrow \infty$.
\end{theorem}

Our main results for second price auctions show an \emph{anytime} convergence for each bidder: as long as $T_0$ is large enough, each bidder will bid truthfully at any time $t$, with high probability. The main technical contribution in this paper is the proof for Theorem~\ref{thm:convergence-spa}. 

\subsection{Proof of Theorem~\ref{thm:convergence-spa}}\label{sec:proof-convergence-spa}
In this section, we summarize the proof of Theorem~\ref{thm:convergence-spa}. To show that, we propose the following lemmas, in which the complete proofs are deferred to \Appendix{missing_proofs}.

Firstly, we characterize that in the pure exploration phase, each bidder gains significantly greater utility when bidding truthfully.
\begin{lemma}\label{lem:excess-utility-single-round-exploration}
For any fixed value $v$, any bid $b\neq v$ and any time $t\leq T_0$, we have for each bidder $i$,
\begin{eqnarray*}
\PP\left(u_{i,t}((v, b_{-i,t}); v) - u_{i,t}((b, b_{-i, t}); v) \geq \frac{1}{H}\right) \geq \frac{\tau}{n}
\end{eqnarray*}
\end{lemma}

Then by a standard Chernoff bound, we can argue the accumulative utility advantage obtained by bidding truthfully will be large enough, for any time $t$ in exploration phase.
\begin{lemma}\label{lem:excess-utility-multiple-rounds-exploration}
For any fixed $v$, any bid $b \neq v$ and any time $t\leq T_0$, we have for each bidder $i$, 
\begin{eqnarray*}
\sum_{s\leq t} u_{i,s}((v, b_{-i, s}); v) - u_{i,s}((b, b_{-i,s}); v) \geq \frac{\tau t}{2nH}
\end{eqnarray*}
holds with probability at least $1-\exp\left(-\frac{\tau^2 t}{2n^2H^{2}}\right)$.
\end{lemma}

Finally, we show that if the cumulative utility advantage of truthful bidding is large enough to satisfy requirements of the mean-based learning algorithms, then truthful bidding still gains significant greater utility for each bidder at time $t > T_0$.
\begin{lemma}\label{lem:excess-utility-single-round-exploitation}
For any $t > T_0$, suppose $\sum_{s\leq t} u_{i,s}((v, b_{-i, s}); v) - u_{i,s}((b, b_{-i,s}); v) \geq \gamma_t t$ holds for any fixed $v, b\neq v$ and each bidder $i$, then
\begin{eqnarray*}
u_{i,t+1}((v, b_{-i, t+1}); v) - u_{i,t+1}((b, b_{-i,t+1}); v) \geq \frac{1}{H}
\end{eqnarray*}
holds with probability at least $\frac{\tau}{2n}$, for any fixed value $v$, bid $b\neq v$ and each bidder $i$.
\end{lemma}

Given the above three auxiliary lemmas, we prove Theorem~\ref{thm:convergence-spa} in the following.  
\begin{proof}[Proof of Theorem~\ref{thm:convergence-spa}]
One of the key techniques used in this paper is the partitioning of the time steps into buckets with a geometric partitioning scheme. In particular, we divide time steps $t > T_0$ to several episodes as follows, $\Gamma_1 = [T_0+1, T_1], \Gamma_2=[T_1+1, T_2],...$, such that $\forall k\geq 1$,  $T_k = \left\lfloor \frac{\tau T_{k-1}}{4\gamma_{T_{k}}nH}\right\rfloor$. We always choose the smallest $T_k$ to satisfy this condition.\footnote{$T_k$ always exists since $\gamma_t t \rightarrow \infty$ as $t \rightarrow \infty$.} The total time steps of each episode $|\Gamma_k| = T_k - T_{k-1}, \forall k\geq 1$.
Then we show the following claim, which states that in each time bucket the expected utility doesn't deviate too much. 

\begin{claim}\label{claim:convergence-spa}
Let event $\mathcal{E}_k$ be $\sum_{s\leq T_k} u_{i,s}((v, b_{-i, s}); v) - u_{i,s}((b, b_{-i,s}); v) \geq \frac{\tau T_k}{4nH}$ holds for all $i$, given any fixed $v, b\neq v$.
Then the event $\mathcal{E}_k$ holds with probability at least $1 - \sum_{\ell=0}^k \exp\left(-\frac{|\Gamma_\ell|\tau^2}{32n^2H^2}\right)$.
\end{claim}

We prove the above claim by induction. If $k=0$, the claim holds by Lemma~\ref{lem:excess-utility-multiple-rounds-exploration}. We assume the claim holds for $k$, then we argue the claim still holds for $k+1$. We consider any time $t\in \Gamma_{k+1}$, given event $\mathcal{E}_k$ holds, we have 
\begin{eqnarray}\label{eq:mean-based-in-episode}
&\sum_{s\leq t} u_{i,s}((v, b_{-i, s}); v) - u_{i,s}((b, b_{-i,s}); v)  \notag \\
&\geq \sum_{s\leq T_k} u_{i,s}((v, b_{-i, s}); v) - u_{i,s}((b, b_{-i,s}); v) \notag \\
&\geq \frac{\tau T_k}{4nH} \geq \gamma_t t,
\end{eqnarray}
where the first inequality is based on the fact that truth-telling is the dominant strategy in second price auctions, the second inequality holds because of the induction assumption and the last inequality hold because 
\begin{eqnarray*}
\gamma_t t\leq \gamma_{T_{k+1}} T_{k+1}  =  \gamma_{T_{k+1}}\cdot \left\lfloor \frac{\tau T_{k}}{4\gamma_{T_{k+1}}nH}\right\rfloor \\
\leq \frac{\tau T_k}{4nH}, \forall t\in \Gamma_{k+1}.
\end{eqnarray*}

Then by Lemma~\ref{lem:excess-utility-single-round-exploitation}, given $\mathcal{E}_k$ holds, for any $t\in \Gamma_{k+1}$ we have,
$$\PP\left(u_{i,t}((v, b_{-i, t}); v) - u_{i,t}((b, b_{-i,t}); v) \geq \frac{1}{H}\Big|\mathcal{E}_k\right)\geq \frac{\tau}{2n}
$$

Thus, $\E\left[u_{i,t}((v, b_{-i, t}); v) - u_{i,t}((b, b_{-i,t}); v)\Big|\mathcal{E}_k\right] \geq \frac{\tau}{2nH}$ for any $t\in \Gamma_{k+1}$. By Azuma's inequality (for martingales), we have
{\small
\begin{align*}
&\PP\left(\sum_{s\in \Gamma_{k+1}} u_{i,s}((v, b_{-i, s}); v) - u_{i,s}((b, b_{-i,s}); v) \leq \frac{\tau |\Gamma_{k+1}|}{4nH}\Big| \mathcal{E}_k\right)\\
\intertext{Letting $\Delta_s =  u_{i,s}((v, b_{-i, s}); v) - u_{i,s}((b, b_{-i,s}); v) $,}
&\leq \PP\left(\sum_{s\in \Gamma_{k+1}} \Delta_s  \leq \sum_{s\in\Gamma_{k+1}}\E\left[\Delta_s\Big|\mathcal{E}_k\right]- \frac{\tau |\Gamma_{k+1}|}{4nH}\Big| \mathcal{E}_k\right)\\
&\leq \exp\left(-\frac{|\Gamma_{k+1}|\tau^2}{32n^2H^2}\right)
\end{align*}
}
Therefore, the event $\mathcal{E}_{k+1}$ holds with probability at least $$\left(1-e^{-\frac{|\Gamma_{k+1}|\tau^2}{32n^2H^2}}\right) \cdot \PP(\mathcal{E}_k) \geq  1- \sum_{\ell=0}^{k+1} \exp\left(-\frac{|\Gamma_\ell|\tau^2}{32n^2H^2}\right),$$
which completes the induction and verifies the correctness of Claim~\ref{claim:convergence-spa}. Given Claim~\ref{claim:convergence-spa}, we have the following argument,

For any time $t > T_0$, there exists $k(t)$, s.t., $t\in \Gamma_{k(t)}$, if the event $\mathcal{E}_{k(t)}$ happens, the bidder $i$ will report truthfully with probability at least $1-H\gamma_t$, by the definition of $\gamma_t$-mean-based learning algorithms and the same argument as Eq.~(\ref{eq:mean-based-in-episode}).
Therefore, at any time $t > T_0$, each bidder $i$ will report truthfully with probability at least 
\begin{eqnarray*}
&&1-H\gamma_t - \sum_{\ell=0}^{k(t)} \exp\left(-\frac{|\Gamma_\ell|\tau^2}{32n^2H^2}\right)\\
\end{eqnarray*}

Then we bound $\sum_{\ell=0}^{k(t)} \exp\left(-\frac{|\Gamma_\ell|\tau^2}{32n^2H^2}\right)$ through the following claim, where the proof is deferred to \Appendix{missing_proofs}.

\begin{claim}\label{claim:constant-lower-bound}
Given $\gamma_t \leq \frac{\tau}{8nH}, \forall t > T_0$, $\sum_{\ell=0}^{k(t)} \exp\left(-\frac{|\Gamma_\ell|\tau^2}{32n^2H^2}\right) \leq 4\exp\left(-\frac{\tau^2 T_0}{32n^2H^2}\right)$, when $T_0$ is large enough s.t. $\exp\left(-\frac{\tau^2 T_0}{32n^2H^2}\right) \leq \frac{1}{2}$.
\end{claim}
Combining the above claim, we complete the proof for Theorem~\ref{thm:convergence-spa}.
\end{proof}

\section{Generalizations to other auctions}
\AppendixName{generalize}

In this section, we generalize our results further to first price auctions when each bidder's valuation is drawn uniformly from $V$, and multi-position auctions when we run the Vickrey-Clarke-Groves (VCG) mechanism. For both cases, we break ties randomly. 

\subsection{First Price Auctions}

In first price auctions, the highest bidder wins and pays her bid. Therefore the utility function of each bidder $i$ at time $t$ can be defined as,
\begin{eqnarray}\label{eq:utility-fpa}
u_{i,t}((b, b_{-i, t}); v) = (v - b)\cdot\1\{b \geq m_{i, t}\}
\end{eqnarray}

It is well-known, first price auctions are not truthful and each bidder will underbid her value to manipulate the auctions. Bayesian Nash Equilibrium (BNE) bidding strategy is hard to characterize in general first price auctions. %
Only the BNE bidding strategy for i.i.d bidders in first price auctions, is fully understood, e.g.~\cite{krishna-book}.

In this paper, for simplicity, we focus on the setting that there are two i.i.d bidders with uniform value distribution over $V$. The BNE bidding strategy for each bidder is $b=\frac{v}{2}$ when the value is $v$~\cite{krishna-book}, if the value space $V = [0, 1]$. In this work, we assume $V = \left\{\frac{1}{H}, \frac{2}{H},\cdots, 1\right\}$ and we break ties randomly. Under this case, we show each mean-based bidder will converge to play near-BNE bidding strategy point-wisely if the number of initial exploration steps $T_0$ is large enough in the following.

\begin{theorem}\label{thm:convergence-fpa}
Suppose there are two bidders and each bidder's value is i.i.d drawn uniformly from $V$ at random, $H$ is a even positive number, and $T_0$ is large enough, such that $\gamma_t \leq \frac{1}{4H^3}, \forall t \geq T_0$. Then at time $t > T_0$, each $\gamma_t$-mean-based learner $i$ in repeated first price auctions will bid $b_t$, s.t. $\frac{v}{2} \leq b_t < \frac{v}{2} + \frac{1}{H}$ with probability at least $1 - H\gamma_t - \exp\left(-\frac{(H-1)T_0}{32(4H^3 + 1)H^4}\right) \frac{\log t}{\log\left(\frac{4H^3 + H}{4H^3 + 1}\right)}$, for any fixed $v_{i, t}$.
\end{theorem}

The proof is significantly different than the proof of second price auctions. Firstly, random tie-breaking makes the analysis more difficult in first price auctions compared with the one in second price auctions. Secondly, $b=\lceil \frac{v}{2} \rceil$ \footnote{$\lceil x \rceil$ means rounding $x$ up to the nearest point in $V$.} is not a dominant bidding strategy, we need a more complex time-splitting scheme to make induction procedure works in first price auctions. We defer the proof to Appendix~\ref{app:convergence-fpa}. We believe our proof for firs price auctions is general enough and can be extended to handle more than two bidders setting.
 
\subsection{Multi-Position VCG Auctions}

In multi-position auctions, there are $k$  positions where $k < n$ and position multipliers 
$p_1 \geq \dots \geq p_k \geq 0 = p_{k+1} = \dots = p_{n}$ (position multipliers determine the relative values or click-through rates of the different positions). In particular, we will also say that 
$p_i-p_{i+1} \geq \rho$ for all $i \leq k$. In this paper, we run VCG mechanism for multi-position auctions. For more details of multi-position VCG auctions, see Appendix~\ref{app:missing-details}.

Denote $z^{(k)}_{i}$ be the $k$-th largest value from the bidders other than bidder $i$, $G^{(k)}_i$ be the (CDF) distribution of $z^{(k)}_{i}$, and $g^{(k)}_{i}$ be the associated PDF. We propose the following thickness assumption for distribution $G^{(k)}_i$, for our theoretical purpose.

\begin{assumption}[Thickness Assumption of $G^{(k)}_i$]\label{assumption:thickness-assumption-multi-position}
There exists a constant $\tau > 0$ (may depend on $n$), s.t., $g_i^{(k)}(v) \geq \tau, \forall i\in [n], \forall v \in V$. Without loss of generality, we assume $\tau \leq \frac{1}{H^{n-1}}$.
\end{assumption}

It is well-known that the VCG auction is truthful. Applying the same technique, we can show the truthful bid has an advantage compared with all the other bids in the exploration phase, as well as in the mean-based exploitation phase. Similarly, we show the following convergence results of mean-based bidders in multi-position VCG auctions.

\begin{theorem}\label{thm:convergence-multi-position}
Suppose assumption~\ref{assumption:thickness-assumption-multi-position} holds and $T_0$ is large enough, such that $\exp\left(-\frac{\tau^2 \rho^2 T_0}{32n^2H^2}\right) \leq \frac{1}{2}$ and $\gamma_t \leq \frac{\tau\rho}{8nH}, \forall t \geq T_0$. Then at time $t > T_0$, each $\gamma_t$-mean-based learner $i$ will $b_t = v_{i, t}$ in repeated multi-position VCG auctions with probability at least $1 - H\gamma_t - 4\exp\left(-\frac{\tau^2 \rho^2 T_0}{32n^2H^2}\right)$, for any fixed $v_{i, t}$.
\end{theorem}

\section{Experiments}
\ggnote{Connect it with previous section. Talk about results in terms of contextual bandits.}

\comment{
\begin{figure*}[h]
\centering
\begin{subfigure}{0.246\textwidth}
\centering
\includegraphics[width=1.\textwidth]{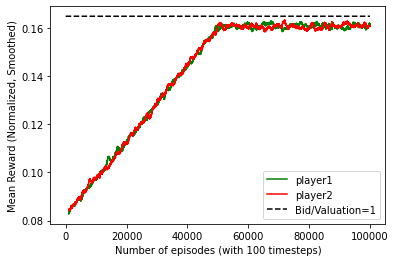}
\end{subfigure}
\begin{subfigure}{0.246\textwidth}
\centering
\includegraphics[width=1.\textwidth]{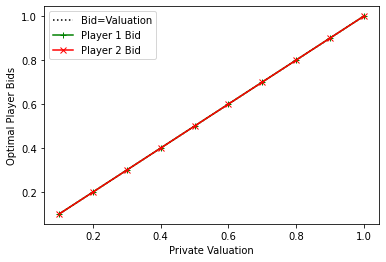}
\end{subfigure}
\begin{subfigure}{0.246\textwidth}
\centering
\includegraphics[width=1.\textwidth]{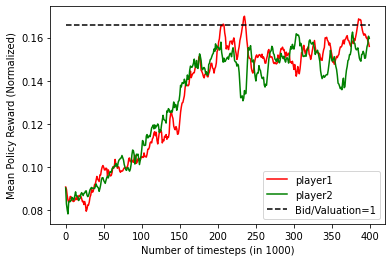}
\end{subfigure}
\caption{Second Price Auctions}
\label{fig:spa}
\vspace{-10pt}
\end{figure*}
}

In this section, we describe the experiments using Contextual Mean-Based Algorithms and  Deep-Q Learning agents participating in repeated first price and second price auctions. In these repeated auctions, the private valuations of both players are sampled independently from identical uniform distributions. The observation for each agent is defined by its private valuation and its reward by the auction outcome (with ties broken randomly).

In the first set of experiments, we study the convergence of two independent learning agents following an $\eps$-Greedy policy in first and second price auction. We use the setting of $H=10$ wherein both agents only observe their private valuation and the respective reward as an auction outcome.

In both cases, we observe the bidders converge to the BNE after several time steps.
There is a slight gap between the (observed) mean reward and utility under BNE as the value of $\eps$ (randomly exploration probability) has a floor of $0.05$. We also observe
that in the exploitation phase, the bidding converges completely to the BNE in the contextual bandit setting, which exactly matches our theory in Figures~\ref{fig:contextual-spa} and~\ref{fig:contextual-fpa}.
We also include some experiments when the number of agents is larger than $2$ in \Appendix{more_experiments} and we
find that the it still converges to BNE.

    \subsection{Extensions to Deep-Q Learning}
    Contextual Mean Based Algorithms are a broad class of algorithms but can be very expensive to implement  if we run a new instance for each possible value and the number of values are large.
    In line with modern machine learning,  one way to mitigate this in practice is to augment it via Deep Q-Learning. To be more concrete, we model the learner by using a deep network with input as the private value and ask it to choose one of many bids. We model this as a reinforcement learning problem where the agents state is input into a deep neural network. The agent's rewards are then observed over time with a chosen discount rate. The details of Deep Q-Learning model and the set 
    of hyperparameters used to train the two Q models are outlined in \Appendix{more_experiments}.

\begin{figure}[t]
\centering
\begin{subfigure}[b]{0.49\textwidth}
\centering
\includegraphics[width=0.49\linewidth]{dqn/contextual_bandits/spa_reward_epsilon_bandits_10H.png}
\includegraphics[width=0.49\linewidth]{dqn/contextual_bandits/spa_bids_bandits_10H_revised.png}

\caption{Training curve of mean reward of each bidder (left) and roll-out bidding strategy of each bidder (right) in the exploitation phase of contextual $\eps$-Greedy algorithm in second price auctions.}
\label{fig:contextual-spa} 
\end{subfigure}
\begin{subfigure}[b]{0.49\textwidth}
\centering
\includegraphics[width=0.49\linewidth]{dqn/spa_reward_convergence.png}
\includegraphics[width=0.49\linewidth]{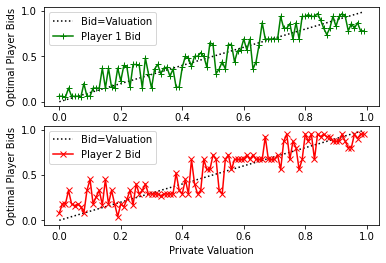}
\caption{Training curve of mean reward of each bidder (left) and roll-out bidding strategy of each bidder (right) in the exploitation phase of Deep-Q Learning algorithm in second price auctions.}
\label{fig:dqn-spa}
\end{subfigure}
\caption{Simulation results for second price auctions}
\label{fig:spa}
\vspace{-10pt}
\end{figure}

\begin{figure}[t]
\centering
\begin{subfigure}[b]{0.49\textwidth}
\centering
  \includegraphics[clip,width=0.5\columnwidth]{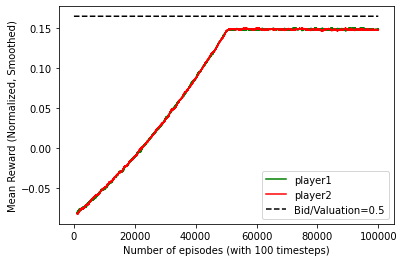}%
  \includegraphics[clip,width=0.5\columnwidth]{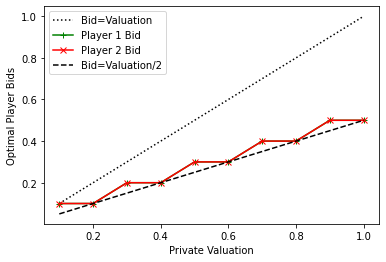}%
\caption{Training curve of mean reward of each bidder (left) and roll-out bidding strategy of each bidder (right) in the exploitation phase of contextual $\eps$-Greedy algorithm in first price auctions.}

\label{fig:contextual-fpa} 
\end{subfigure}
\begin{subfigure}[b]{0.49\textwidth}
\centering
\includegraphics[clip,width=0.5\columnwidth]{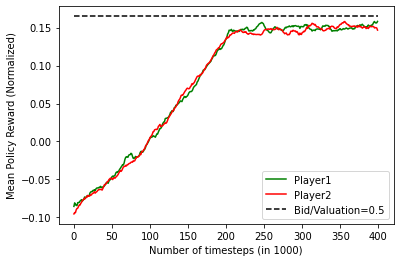}%
\includegraphics[clip,width=0.5\columnwidth]{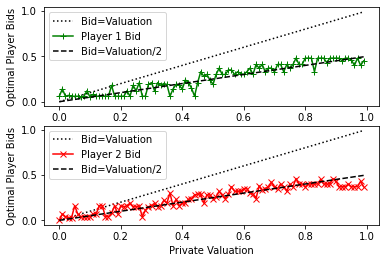}%
\caption{Training curve of mean reward of each bidder (left) and roll-out bidding strategy of each bidder (right) in the exploitation phase of Deep-Q Learning algorithm in first price auctions.}
\label{fig:dqn-fpa}
\end{subfigure}
\caption{Simulation Results for first price auctions}
\label{fig:fpa}
\vspace{-10pt}
\end{figure}

    We use the setting 
    of $H=100$ and consider the observation of the agent as its private valuation.  Again, 
    we observe that both agents converge to BNE, shown in Figures~\ref{fig:dqn-spa} and~\ref{fig:dqn-fpa}. We also study the model 
    with a wider set of states including previously chosen bids and  empirically observe the convergence 
    of independent DQN agents to BNE for both auctions (discussed further in \Appendix{more_experiments}).

\comment{
\begin{figure*}[h]
\centering
\begin{subfigure}{0.246\textwidth}
\centering
\includegraphics[width=1.\textwidth]{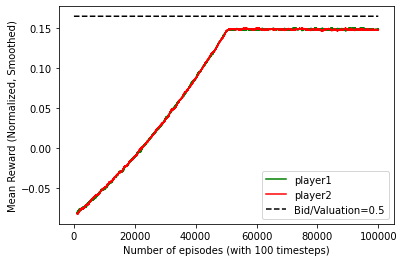}
\end{subfigure}
\begin{subfigure}{0.246\textwidth}
\centering
\includegraphics[width=1.\textwidth]{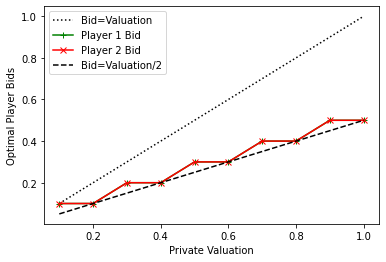}
\end{subfigure}
\begin{subfigure}{0.246\textwidth}
\centering
\includegraphics[width=1.\textwidth]{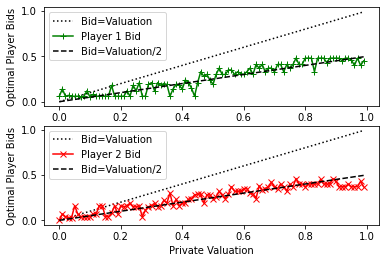}
\end{subfigure}
\begin{subfigure}{0.246\textwidth}
\centering
\includegraphics[width=1.\textwidth]{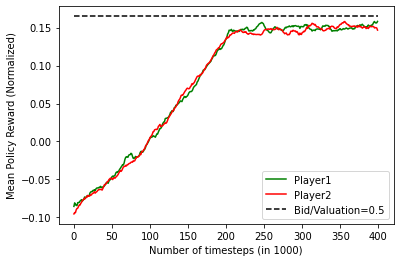}
\end{subfigure}
\caption{First Price Auctions}
\label{fig:fpa}
\vspace{-10pt}
\end{figure*}
}

\bibliographystyle{ACM-Reference-Format}
\bibliography{refs}

\onecolumn
\clearpage
\appendix
\begin{center}
{
\Large
\textbf{
Appendix}
}
\end{center}
\section{Missing Details}\label{app:missing-details}

\subsection{Multi-Position VCG Auctions}
Without loss of generality, we assume that bidders are arranged in descending order of bids and there are no ties, i.e., ~$b_1 > b_2 > \dots > b_n$. 
Since we run VCG auctions, bidder $i$ gets the position $i$ and the payment extracted from the bidder
$i$ is exactly $\sum_{j=i}^m (p_j-p_{j+1}) b_{j}$. The utility that bidder $i$
gets at being in position $i$ is exactly, $\sum_{j=i}^m (p_j - p_{j+1})\cdot (v_i - b_{j})$. When there are ties, we break ties randomly.

\section{Missing Proofs}
\AppendixName{missing_proofs}

\subsection{Proof of Lemma~\ref{lem:excess-utility-single-round-exploration}}
\begin{proof}
We prove this lemma by considering the following cases,
\begin{itemize}
\item If $b \geq v+\frac{1}{H}$, then $m_{i, t} = b$ implies $u_{i,t}((v, b_{-i, t}); v) =  0$, whereas, $u_{i,t}((b, b_{-i, t}); v) = v - b \leq -\frac{1}{H}$ with probability at least $1/n$, because of random tie-breaking. Therefore,
\begin{eqnarray*}
&&\PP\left(u_{i,t}((v, b_{-i, t}); v) - u_{i,t}((b, b_{-i, t}); v)\geq \frac{1}{H}\Big|b\geq v+\frac{1}{H}\right)\\
&\geq& \PP\left(u_{i,t}((v, b_{-i, t}); v) - u_{i,t}((b, b_{-i, t}); v) \geq \frac{1}{H}, m_{i, t} = b\Big|b\geq v+\frac{1}{H}\right)\\
&\geq& \PP\left(u_{i,t}((b, b_{-i, t}); v) \leq -\frac{1}{H}, m_{i, t} = b\Big|b\geq v+\frac{1}{H}\right)\\
&\geq& \frac{1}{n} \left(\frac{1}{H}\right)^{n-1} \geq \frac{\tau}{n}
\end{eqnarray*}
where the second last inequality holds because $\PP\left(m_{i,t} = b\right) \geq \left(\frac{1}{H}\right)^{n-1}$ and the last inequality is based on the fact that $\frac{1}{H^{n-1}} \geq \tau$.
\item If $b \leq v - \frac{1}{H}$, then $m_{i,t} = b$ implies $u_{i, t}((v, b_{-i, t}); v) = v - m_{i,t}\geq \frac{1}{H}$, whereas, $u_{i,t}(b, b_{-i, t}); v)) = 0$ with both probability at least $1/n$ (random tie-breaking). Therefore,
\begin{eqnarray*}
&&\PP\left(u_{i,t}((v, b_{-i, t}); v) - u_{i,t}((b, b_{-i, t}); v)\geq \frac{1}{H}\Big|b\leq v-\frac{1}{H}\right)\\
&\geq& \PP\left(u_{i,t}((v, b_{-i, t}); v) - u_{i,t}((b, b_{-i, t}); v) \geq \frac{1}{H}, m_{i, t} = b\Big|b\leq v-\frac{1}{H}\right)\\
&\geq& \PP\left(u_{i,t}((b, b_{-i, t}); v) = 0, m_{i, t} = b\Big|b\leq v-\frac{1}{H}\right)\\
&\geq& \frac{1}{n} \left(\frac{1}{H}\right)^{n-1} \geq \frac{\tau}{n}
\end{eqnarray*}
\end{itemize}
\end{proof}

\subsection{Proof of Lemma~\ref{lem:excess-utility-multiple-rounds-exploration}}
\begin{proof}
By Lemma~\ref{lem:excess-utility-single-round-exploration}, we have $\E[u_{i,s}((v, b_{-i,s}); v) - u_{i,s}((b, b_{-i,s}); v)] \geq \frac{\tau}{nH}$, for any fixed $v$, $b\neq v$ and $s\leq t$. Then by Chernoff bound, we have
\begin{eqnarray*}
&&\PP\left(\sum_{s\leq t} u_{i,s}((v, b_{-i, s}); v) - u_{i,s}((b, b_{-i,s}); v) \leq \frac{\tau t}{nH}  - \frac{\tau t}{2nH}\right)\\
&\leq& \PP\left(\sum_{s\leq t} u_{i,s}((v, b_{-i, s}); v) - u_{i,s}((b, b_{-i,s}); v) \leq t \E[u_{i,s}((v, b_{-i,s}); v) - u_{i,s}((v, b_{-i,s}); v)] -  \frac{\tau t}{2nH}\right)\\
&\leq& \exp\left(-\frac{2\tau^2 t}{4n^2H^{2}}\right)
\end{eqnarray*}
\end{proof}

\subsection{Proof of Lemma~\ref{lem:excess-utility-single-round-exploitation}}
\begin{proof}
By definition of $\gamma_t$-mean-based learning algorithm, given a value $v$, each bid $b \neq v$ will be selected with probability at most $\gamma_t$ for each bidder $i$, thus, $b_{i,t+1} \neq v_{i, t+1}$ holds with probability at most $H\gamma_t$ for any bidder $i$. By union bound, $m_{i, t+1} \neq z_{i, t+1}$ holds with probability at most $(n-1)H\gamma_t$, for all $i$.
\begin{eqnarray*}
&&\PP\left(u_{i,t+1}((v, b_{-i, t+1}); v) - u_{i,t}((b, b_{-i,t+1}); v) \geq \frac{1}{H}\right)\\
&\geq& \PP\left(u_{i,t+1}((v, b_{-i, t+1}); v) - u_{i,t+1}((b, b_{-i,t+1}); v) \geq \frac{1}{H}, m_{i, t+1} = z_{i, t+1}\right)\\
&\geq& \PP\left(u_{i,t+1}((v, b_{-i, t+1}); v) - u_{i,t+1}((b, b_{-i,t+1}); v) \geq \frac{1}{H}, m_{i, t+1} = z_{i, t+1}, z_{i, t+1} = b\right)\\
&\geq& \left(1 - (n-1)H\gamma_t\right)\cdot \frac{\tau}{n} \geq \frac{\tau}{2n},
\end{eqnarray*}
where the third inequality is based on the same argument in Lemma~\ref{lem:excess-utility-single-round-exploration} and the last inequality holds because $\gamma_t \leq \frac{1}{2nH}$.
\end{proof}

\subsection{Proof of Claim~\ref{claim:constant-lower-bound}}
\begin{proof}
Based on the %
construct of $T_k$ and $\gamma_t \leq \frac{\tau}{8nH}, \forall t > T_0$, we have $T_k \geq 2T_{k-1}$. Then $|\Gamma_\ell| = T_{\ell} - T_{\ell-1} \geq 2^{\ell-1} T_0, \forall \ell\geq 1$, we have
\begin{eqnarray*}
\sum\limits_{\ell=0}^{k(t)}\exp\left(-\frac{|\Gamma_\ell|\tau^2}{32n^2H^2}\right) &\leq& 2\exp\left(-\frac{\tau^2 T_0}{32n^2H^2}\right)+ \sum\limits_{\ell=2}\exp\left(-\frac{2^{\ell-1}\tau^2 T_0}{32n^2H^2}\right)\\
&\leq&  2\exp\left(-\frac{\tau^2 T_0}{32n^2H^2}\right) + \sum_{\ell=1} \exp\left(-\frac{2^{\ell} \tau^2 T_0}{32n^2H^2}\right)\\
&=&  \exp\left(-\frac{\tau^2 T_0}{32n^2H^2}\right)\cdot \left(2 + \sum_{\ell=1} \exp\left(-\frac{(2^{\ell}-1) \tau^2 T_0}{32n^2H^2}\right)\right)\\
&\leq&  \exp\left(-\frac{\tau^2 T_0}{32n^2H^2}\right)\cdot \left(2 + \sum_{\ell=1} \exp\left(-\frac{\ell \tau^2 T_0}{32n^2H^2}\right)\right)\\
&&(\text{Because } 2^\ell - 1 \geq \ell, \forall \ell\geq 1)\\
&\leq &\exp\left(-\frac{\tau^2 T_0}{32n^2H^2}\right)\cdot \left(2 + \frac{1}{1 - \exp\left(-\frac{\tau^2 T_0}{32n^2H^2}\right)}\right)\\
&\leq& 4\exp\left(-\frac{\tau^2 T_0}{32n^2H^2}\right),
\end{eqnarray*}
where the last inequality holds, because $\exp\left(-\frac{\tau^2 T_0}{32n^2H^2}\right) \leq \frac{1}{2}$ if $T_0$ is large enough.
\end{proof}

\subsection{Proof of Theorem~\ref{thm:convergence-fpa}}\label{app:convergence-fpa}

Here we slightly abuse the notation, let $\lceil \frac{v}{2}\rceil := \{b\in V: b\geq \frac{v}{2}, \text{ and } b \leq\frac{v}{2}+\frac{1}{H}\}$. Notice, if $\frac{v}{2}\in V$, $\lceil \frac{v}{2}\rceil=\frac{v}{2}$. If $\frac{v}{2} \notin V$, $\lceil \frac{v}{2}\rceil=\frac{v}{2} + \frac{1}{2H}$. We prove this theorem based on the following claims,

\begin{claim}\label{claim:fpa-1}
For any $t \leq T_0$, any fixed $v$, any bid $b\neq \lceil \frac{v}{2}\rceil$, for each bidder $i$ we have
\[\expects{b_{-i,t}}{u_{i,t}((\lceil \frac{v}{2}\rceil, b_{-i,t}); v) - u_{i,t}((b,b_{-i,t}); v)} \geq \frac{1}{2H^2}\]
\end{claim}
\begin{proof}
Let $\mathcal{U}_V$ denote the uniform distribution on $V$.
Note, we assume the random tie-breaking in this paper, then we can rewrite the expected utility of bidder $i$ when $t \leq T_0$ in the following way
\begin{eqnarray*}
\expects{b_{-i,t}\sim \mathcal{U}_{V}}{u_{i,t}((b, b_{-i,t}); v)} = (v - b) \cdot \left(\left(b-\frac{1}{H}\right) + \frac{1}{2}\cdot \frac{1}{H}\right) = (v- b)\cdot \left(b - \frac{1}{2H}\right)
\end{eqnarray*}

Then we consider two different cases in the following,
\begin{itemize}
\item $\frac{v}{2} \in V$, let $b = \frac{v}{2} + \alpha$, where $\alpha \geq \frac{1}{H}$ or $\alpha \leq -\frac{1}{H}$. Thus, we have
\begin{eqnarray*}
&&\left(v - \frac{v}{2}\right)\cdot \left(\frac{v}{2} - \frac{1}{2H}\right) - (v - b)\cdot \left(b - \frac{1}{2H}\right)\\
&=& \frac{v^2}{4} - \frac{v}{4H} - \left(\frac{v^2}{4} - \alpha^2 -\frac{v}{4H} + \frac{\alpha}{2H}\right)\\
&=& \alpha^2 - \frac{\alpha}{2H} \geq \frac{1}{2H^2}
\end{eqnarray*}

\item $\frac{v}{2}\notin V$, then $\lceil \frac{v}{2}\rceil = \frac{v}{2} + \frac{1}{2H}$. Let $b = \lceil \frac{v}{2}\rceil + \alpha$, where $\alpha \geq \frac{1}{H}$ or $\alpha \leq -\frac{1}{H}$. Thus, we have
\begin{eqnarray*}
&&\left(v - \left\lceil\frac{v}{2}\right\rceil\right)\cdot \left(\left\lceil\frac{v}{2}\right\rceil - \frac{1}{2H}\right) - (v - b)\cdot \left(b - \frac{1}{2H}\right)\\
&=& \left(\frac{v}{2} - \frac{1}{2H}\right)\cdot \frac{v}{2} - \left(\frac{v}{2}-\frac{1}{2H} -\alpha\right)\cdot\left(\frac{v}{2} + \alpha\right)\\
&=& \alpha^2 + \frac{\alpha}{2H} \geq \frac{1}{2H^2}
\end{eqnarray*}
\end{itemize}
Combining the above two cases, we complete the proof.
\end{proof}

\begin{claim}\label{claim:fpa-2}
For any fixed value $v$, any $t\leq T_0$, any bid $b \neq \lceil\frac{v}{2}\rceil$, we have for each bidder $i$,
\begin{eqnarray*}
\PP\left(\sum_{s\leq t} u_{i, s}((\lceil \frac{v}{2}\rceil, b_{-i, s}); v) - \sum_{s\leq t} u_{i, s}((b, b_{-i, s}); v) \leq \frac{\tau t}{2nH}\right) \leq \exp\left(-\frac{t}{8H^4}\right)
\end{eqnarray*}
\end{claim}
\begin{proof}
By Claim~\ref{claim:fpa-1} and Chernoff bound, we have
\begin{eqnarray*}
&&\PP\left(\sum_{s\leq t} u_{i, s}((\lceil \frac{v}{2}\rceil, b_{-i, s}); v) - \sum_{s\leq t} u_{i, s}((b, b_{-i, s}); v) \leq \frac{t}{4H^2}\right) \\
&\leq & \PP\left(\sum_{s\leq t} u_{i, s}((\lceil \frac{v}{2}\rceil, b_{-i, s}); v) - \sum_{s\leq t} u_{i, s}((b, b_{-i, s}); v) \leq \sum_{s\leq t}\expects{b_{-i,s}}{u_{i,s}((\lceil \frac{v}{2}\rceil, b_{-i,s}); v) - u_{i,s}((b,b_{-i,s}); v)} - \frac{t}{4H^2}\right)\\
&\leq& \exp\left(-\frac{2t}{16H^4}\right) = \exp\left(-\frac{t}{8H^4}\right)
\end{eqnarray*}
\end{proof}

\begin{claim}\label{claim:fpa-3}
For any $t > T_0$, for any fixed $v$, any bid $b\neq \lceil \frac{v}{2}\rceil$ and each bidder $i$, suppose $\sum_{s\leq t} u_{i,s}((\lceil \frac{v}{2}\rceil, b_{-i, s}); v) - u_{i,s}((b, b_{-i,s}); v) \geq \gamma_t t$ holds, then for any fixed value $v$, any bid $b\neq \lceil \frac{v}{2}\rceil$ and each bidder $i$, we have,
\begin{eqnarray*}
\E_{b_{-i, t+1}}\left[u_{i,t+1}((\lceil \frac{v}{2}\rceil, b_{-i, t+1}); v) - u_{i,t+1}((b, b_{-i,t+1}); v)\right] \geq \frac{1}{2H^2}
\end{eqnarray*}
\end{claim}

\begin{proof}
We assume for each bidder $i$, with probability $\eta_t^i$ bids $b_{i, i+1} = \lceil \frac{v_{i, t+1}}{2}\rceil$ and with probability $1-\eta_t^i$ bids $b_{i, t+1}\neq \lceil \frac{v_{i, t+1}}{2}\rceil$. By the condition that $\sum_{s\leq t} u_{i,s}((\lceil \frac{v}{2}\rceil, b_{-i, s}); v) - u_{i,s}((b, b_{-i,s}); v) \geq \gamma_t t$ and definition of mean-based learning algorithm, we have $\eta_t^i \geq 1 - H\gamma_t$ for each bidder $i$.

Let $b_{j, t+1}, v_{j, t+1}$ be the bid and value from the other bidder $j\neq i$ at time $t+1$, respectively. Then we can show the lower bound of the expected utility for bidder $i$, when bid $b \leq \frac{1}{2}$, in the following, 

\begin{eqnarray*}
&&E_{b_{-i, t+1}}\left[u_{i,t+1}((b, b_{-i,t+1}); v)\right]\\
&=& (v - b) \cdot\left(\PP(b > b_{j, t+1}) + \frac{1}{2} \cdot \PP(b_{j, t+1} = b)\right)\\
&\geq & \eta_t^j \cdot (v-b)\cdot \left(\PP(b > \lceil \frac{v_{j, t+1}}{2}\rceil) + \frac{1}{2} \cdot \PP(\lceil \frac{v_{j, t+1}}{2}\rceil = b)\right) 
\end{eqnarray*}

For any $b \leq \frac{1}{2}$, $2b\in V$. Then if $v_{j, t+1} \leq 2b - \frac{2}{H}$, $\lceil \frac{v_{j, t+1}}{2}\rceil < b$. Therefore, $\PP(b > \lceil \frac{v_{j, t+1}}{2}\rceil) = 2b - \frac{2}{H}$. Notice, when $v_{j, t+1} = 2b - \frac{2}{H}$ or $v_{j, t+1} = 2b - \frac{1}{H}$, $\lceil \frac{v_{j, t+1}}{2}\rceil  = b$, then $\PP\left(\left\lceil \frac{v_{j, t+1}}{2}\right\rceil = b\right) = \frac{2}{H}$. Therefore, we can lower bound the expected utility for bidder $i$, when bid $b \leq \frac{1}{2}$,
\begin{eqnarray*}
E_{b_{-i, t+1}}\left[u_{i,t+1}((b, b_{-i,t+1}); v)\right] \geq 2\eta_t^j\cdot (v-b)\cdot\left(b - \frac{1}{2H}\right)
\end{eqnarray*}

Similarly, we can upper bound the expected utility for bidder $i$, when bid $b \leq \frac{1}{2}$, shown as below,
\begin{eqnarray*}
&&E_{b_{-i, t+1}}\left[u_{i,t+1}((b, b_{-i,t+1}); v)\right]\\
&\leq & \eta_t^j \cdot (v-b)\cdot \left(\PP(b > \lceil \frac{v_{j, t+1}}{2}\rceil) + \frac{1}{2} \cdot \PP(\lceil \frac{v_{j, t+1}}{2}\rceil = b)\right)  + (1-\eta_t^j)\\
&=& 2\eta_t^j\cdot (v-b)\cdot\left(b - \frac{1}{2H}\right) + (1-\eta_t^j)
\end{eqnarray*}

Combining the above lower bound and upper bound of the expected utility of bidder $i$, we have for any $b \leq \frac{1}{2}$,
\begin{eqnarray*}
&&\E_{b_{-i, t+1}}\left[u_{i,t+1}((\lceil \frac{v}{2}\rceil, b_{-i, t+1}); v) - u_{i,t+1}((b, b_{-i,t+1}); v)\right]\\
&\geq& 2\eta_t^j\cdot  \left((v-\lceil \frac{v}{2}\rceil)\cdot \left(\lceil \frac{v}{2}\rceil- \frac{1}{2H}\right) - (v-b)\cdot \left(b- \frac{1}{2H}\right)\right) - (1-\eta_t^j)\\
&\geq & 2(1-H\gamma_t) \frac{1}{2H^2} - H\gamma_t = (1-H\gamma_t)\frac{1}{H^2} - H\gamma_t,
\end{eqnarray*}
where the last inequality is based on the same argument as in Claim~\ref{claim:fpa-1}. Finally, since $T_0$ is large enough to make $\gamma_t \leq \frac{1}{4H^3}$, then we have $(1-H\gamma_t)\frac{1}{H^2} - H\gamma_t\geq \frac{3}{4}\cdot \frac{1}{H^2} - \frac{1}{4H^2} = \frac{1}{2H^2}$. Thus, for any bid $b \leq \frac{1}{2}$, we have,
\begin{eqnarray*}
\E_{b_{-i, t+1}}\left[u_{i,t+1}((\lceil\frac{v}{2}\rceil, b_{-i, t+1}); v) - u_{i,t+1}((b, b_{-i,t+1}); v)\right] \geq \frac{1}{H}
\end{eqnarray*}
For bid $b \geq \frac{1}{2} + \frac{1}{H}$, $\PP(b > \lceil \frac{v_{j, t+1}}{2}\rceil) = 1$ and $\PP(\lceil \frac{v_{j, t+1}}{2}\rceil = b) = 0$, then it is trivially to show for any $v\in V$, 
\begin{eqnarray*}
\E_{b_{-i, t+1}}\left[u_{i,t+1}((\frac{1}{2}, b_{-i, t+1}); v) - u_{i,t+1}((b, b_{-i,t+1}); v)\right] \geq \frac{1}{H}
\end{eqnarray*}

Combining the case that $b \leq \frac{1}{2}$, we complete the proof.
\end{proof}

Similarly to the proof of Theorem~\ref{thm:convergence-spa}, we divide the time steps $t > T_0$ to several episodes as follows, $\Gamma_1 = [T_0+1, T_1], \Gamma_2=[T_1+1, T_2],...$, such that $\forall k\geq 1$,  $T_k = \left\lfloor \frac{\left(\frac{1}{4H^2} + 1\right)T_{k-1}}{\gamma_{T_{k}}+1}\right\rfloor$. We always choose the smallest $T_k$ to satisfy this condition. This $T_k$ always exists since $(\gamma_t + 1) t \rightarrow \infty$ as $t\rightarrow \infty$. The total time steps of each episode $|\Gamma_k| = T_k - T_{k-1}, \forall k\geq 1$.
Then we show the following claim holds.

\begin{claim}\label{claim:fpa-4}
Let event $\mathcal{E}_k$ be $\sum_{s\leq T_k} u_{i,s}((\lceil \frac{v}{2}\rceil, b_{-i, s}); v) - u_{i,s}((b, b_{-i,s}); v) \geq \frac{T_k}{4H^2}$ holds for all $i$, any fixed $v$, and any bid $b\neq \lceil \frac{v}{2}\rceil$.
Then the event $\mathcal{E}_k$ holds with probability at least $1 - \sum_{\ell=0}^k \exp\left(-\frac{|\Gamma_\ell|}{32H^4}\right)$.
\end{claim}

We prove the above claim by induction. If $k=0$, the claim holds by Claim~\ref{claim:fpa-2}. We assume the claim holds for $k$, then we argue the claim still holds for $k+1$. We consider any time $t\in \Gamma_{k+1}$, given event $\mathcal{E}_k$ holds, we have 
\begin{equation}\label{eq:mean-based-in-episode}
\begin{aligned}
\sum_{s\leq t} u_{i,s}((\lceil \frac{v}{2}\rceil, b_{-i, s}); v) - u_{i,s}((b, b_{-i,s}); v) &\geq& \sum_{s\leq T_k} u_{i,s}((\lceil \frac{v}{2}\rceil, b_{-i, s}); v) - u_{i,s}((b, b_{-i,s}); v) - (t - T_k)\\
&\geq& \frac{T_k}{4H^2} - T_{k+1} + T_k = \left(\frac{1}{4H^2}+1\right)T_k - T_{k+1} \geq \gamma_t t,
\end{aligned}
\end{equation}
where the first inequality holds because $u_{i,s}((\lceil \frac{v}{2}\rceil, b_{-i, s}); v) - u_{i,s}((b, b_{-i,s}); v) \geq -1, \forall s > T_k$ and the final inequality holds because of the induction assumption and the last inequality hold because 
\begin{eqnarray*}
\gamma_t t + T_{k+1}\leq (\gamma_{T_{k+1}} + 1) T_{k+1} =  (\gamma_{T_{k+1}}+1)\cdot \left\lfloor \frac{\left(\frac{1}{4H^2} + 1\right)T_{k}}{\gamma_{T_{k+1}}+1}\right\rfloor \leq \left(\frac{1}{4H^2} + 1\right)T_{k}, \forall t\in \Gamma_{k+1}.
\end{eqnarray*}

Then by Claim~\ref{claim:fpa-3}, given $\mathcal{E}_k$ holds, for any $t\in \Gamma_{k+1}$ we have, $\E\left[u_{i,t}((\lceil \frac{v}{2}\rceil, b_{-i, t}); v) - u_{i,t}((b, b_{-i,t}); v)\Big|\mathcal{E}_k\right] \geq \frac{1}{2H^2}$ for any $t\in \Gamma_{k+1}$. By Azuma's inequality (for martingale), we have
\begin{eqnarray*}
&&\PP\left(\sum_{s\in \Gamma_{k+1}} u_{i,s}((\lceil \frac{v}{2}\rceil, b_{-i, s}); v) - u_{i,s}((b, b_{-i,s}); v) \leq \frac{|\Gamma_{k+1}|}{4H^2}\Big| \mathcal{E}_k\right)\\
&\leq& \PP\left(\sum_{s\in \Gamma_{k+1}} u_{i,s}((\lceil \frac{v}{2}\rceil, b_{-i, s}); v) - u_{i,s}((b, b_{-i,s}); v) \leq \sum_{s\in\Gamma_{k+1}}\E\left[u_{i,s}((\lceil \frac{v}{2}\rceil, b_{-i, s}); v) - u_{i,s}((b, b_{-i,s}); v)\Big|\mathcal{E}_k\right]- \frac{|\Gamma_{k+1}|}{4H^2}\Big| \mathcal{E}_k\right)\\
&\leq& \exp\left(-\frac{|\Gamma_{k+1}|}{32H^4}\right)
\end{eqnarray*}

Therefore, the event $\mathcal{E}_{k+1}$ holds with probability at least $$\left(1-\exp\left(-\frac{|\Gamma_{k+1}|}{32 H^4}\right)\right) \cdot \PP(\mathcal{E}_k) \geq  1- \sum_{\ell=0}^{k+1} \exp\left(-\frac{|\Gamma_\ell|}{32 H^4}\right),$$
which completes the induction. Given Claim~\ref{claim:convergence-spa}, we have the following argument,

For any time $t > T_0$, there exists $k(t)$, s.t., $t\in \Gamma_{k(t)}$, if the event $\mathcal{E}_{k(t)}$ happens, the bidder $i$ will report $b_t = \lceil \frac{v_{i, t}}{2}\rceil$ at least $1-H\gamma_t$, by the definition of $\gamma_t$-mean-based learning algorithms and the same argument as Eq.~(\ref{eq:mean-based-in-episode}).
Therefore, at any time $t > T_0$, each bidder $i$ will report truthfully with probability at least 
\begin{eqnarray*}
&&1-H\gamma_t - \sum_{\ell=0}^{k(t)} \exp\left(-\frac{|\Gamma_\ell|}{32H^4}\right)\\
&=&1-H\gamma_t- \sum_{\ell=0}^{k(t)} \exp\left(-\frac{|\Gamma_\ell|}{32H^4}\right)\\
\end{eqnarray*}

We then bound $k(t)$. First, $T_k \geq \left(\frac{4H^3 + H}{4H^3+ 1}\right)T_{k-1}$, since $\gamma_t \leq \frac{1}{4H^3}, \forall t > T_0$. Therefore, we have $\left(\frac{4H^3 + H}{4H^3+ 1}\right)^{(k(t)-1)}T_0 \leq t$, which implies, $k(t) + 1\leq 2 + \frac{\log(t/T_0)}{\log\left(\frac{4H^3 + H}{4H^3+ 1}\right)} \leq \frac{\log t}{\log\left(\frac{4H^3 + H}{4H^3+ 1}\right)}$. In addition, we have $|\Gamma_\ell| \geq \frac{H-1}{4H^3+ 1}T_0$.
Combining the above arguments together, we complete the proof.

\noindent\textbf{Remark.} Let $p_{i}(t)$ be the probability that each bidder $i$ bids $\lceil \frac{v_{i, t}}{2} \rceil$ at time $t$, for any fixed $v_{i, t}$. As long as $T_0 = \Omega(\log\log t)$, $p_{i}(t) \rightarrow 1$ as $t \rightarrow \infty$.

\subsection{Proof of Theorem~\ref{thm:convergence-multi-position}}

This proof exactly follows the same technique in Theorem~\ref{thm:convergence-spa}. Here we only mention the difference in multi-position VCG auctions compared with second price auctions, summarized in the following two claims.

\begin{claim}\label{claim:multi-position-1}
For any fixed value $v$, and bid $b\neq v$ and any time $t \leq T_0$, for each bidder $i$, we have
\begin{eqnarray*}
\PP\left(u_{i,t}((v, b_{-i, t}); v) - u_{i, t}(b, b_{-i, t}); v) \geq \frac{\rho}{H}\right) \geq \frac{\tau}{n}
\end{eqnarray*}
\end{claim}

\begin{proof}
Firstly, truthful bidding is the weakly dominant strategy for each bidder $i$, thus, for any $b_{-i, t}$ and $b\neq v$, $u_{i,t}((v, b_{-i, t}); v) - u_{i, t}(b, b_{-i, t}); v) \geq 0$. 

Then we focus on the case that $b \geq v+\frac{1}{H}$ here. It is analogous to show for the case $b \leq v-\frac{1}{H}$ and we omit here. Consider the other bidders all bid $b$, since $m < n$, bidder $i$ wins no slot if she bids truthfully, i.e. $u_{i, t}((v, b_{-i, t}); v) = 0$ if $b_{j, t} = b, \forall j\neq i$. However, if bidder $i$ bids $b$, by random tie-breaking, bidder $i$ will win a slot with probability $1/n$ and pay $b$ if she wins, i.e., $u_{i,t}((b; b_{-i, t}); v) \leq \rho(v - b) \leq -\frac{\rho}{H}$ if bidder $i$ wins a slot.

Notice, the probability that the other bidders all bid $b$ is $\frac{1}{H^{n-1}}$, therefore 
\begin{eqnarray*}
\PP\left(u_{i,t}((v, b_{-i, t}); v) - u_{i, t}(b, b_{-i, t}); v) \geq \frac{\rho}{H}\right) \geq \frac{1}{n H^{n-1}} \geq \frac{\tau}{n}
\end{eqnarray*}
\end{proof}

\begin{claim}\label{claim:multi-position-2}
For any $t > T_0$, suppose $\sum_{s\leq t} u_{i,s}((v, b_{-i, s}); v) - u_{i,s}((b, b_{-i,s}); v) \geq \gamma_t t$ holds for any fixed $v, b\neq v$ and each bidder $i$, then
\begin{eqnarray*}
u_{i,t+1}((v, b_{-i, t+1}); v) - u_{i,t+1}((b, b_{-i,t+1}); v) \geq \frac{\rho}{H}
\end{eqnarray*}
holds with probability at least $\frac{\tau}{2n}$, for any fixed value $v$, bid $b\neq v$ and each bidder $i$.
\end{claim}
\begin{proof}
Similarly to Lemma~\ref{lem:excess-utility-single-round-exploitation}, by definition of $\gamma_t$-mean-based learning algorithm and the condition assumed in the claim, each bidder $i$ will submit bid $b=v_{i,t+1}$ at least $1-H\gamma_t$. Moreover, by Assumption~\ref{assumption:thickness-assumption-multi-position}, the probability that the $k$-th largest value from the other bidders is $b$, is at least $\tau$. Following the same argument in Claim~\ref{claim:multi-position-1}, we have 
\begin{eqnarray*}
u_{i,t+1}((v, b_{-i, t+1}); v) - u_{i,t+1}((b, b_{-i,t+1}); v) \geq \frac{\rho}{H}
\end{eqnarray*}
holds with probability at least $(1-H\gamma_t)^{n-1} \frac{\tau}{n} \geq (1-H(n-1)\gamma_t) \frac{\tau}{n}$.
\end{proof}

Given the above two claims, following the exactly same proof steps as in Theorem~\ref{thm:convergence-spa}, we complete the proof for Theorem~\ref{thm:convergence-multi-position}.

\section{Additional Experiments}
\AppendixName{more_experiments}
In this section, we outline the experimental setup for the contextual bandit experiments
as well as the deep Q-learning experiments. In the contextual bandit setting, we use the state to simply represent the private value of the user and reward to be the utility of the auction. In the RL setup, we define the observations of each agent as their private valuation, the state transitions as the next private valuation sampled randomly, the rewards as the utility of the auction and the actions as the discrete bid chosen at each round. Note that in the RL setup where the state is the private value and the next private valuation is sampled randomly, the proposed setting would be similar to the contextual bandit setting. Our learning algorithm is the $\epsilon$-greedy which is a Mean-Based Algorithm studied earlier. In the experimental setup, we considered a finite horizon episode consisting of $N$ auction rounds (where $N$ is chosen as 100 typically in the experiments).
\subsection{Experiment Details}
\begin{figure}
\centering
\includegraphics[width=0.49\linewidth]{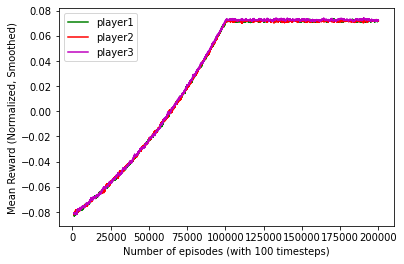}
\includegraphics[width=0.49\linewidth]{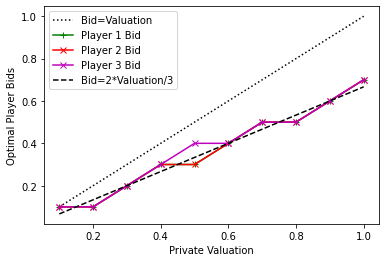}
\caption{Training curve of mean reward of each bidder (left) and roll-out bidding strategy of each bidder (right) in the exploitation phase of contextual $\eps$-Greedy algorithm in first price auctions for three bidders.}
\label{fpa_multi_agent}
\end{figure}

\noindent \textbf{Contextual Bandits Setting:}
In first and second price auctions, we anneal the $\epsilon$ for both players from $1.0$ to $0.05$ over 50,000 episodes with 100 rounds each. To measure the robustness of our results for more than two players, we evaluate the equilibrium performance for three agents participating in a first price auction in Figure~\ref{fpa_multi_agent}.

\noindent \textbf{Deep Q Learning Setting:}
In the DQN experiments mentioned earlier, we used Double Deep Q Networks \cite{van2015deep} with Dueling \cite{wang2016dueling} and Prioritized Experience Replay \cite{schaul2015prioritized} to train the two agents with identical hyperparameters. In the experiments, the Q network is a fully-connected network with hidden dimensions [256, 256] and tanh non-linearity, the number of discrete states $H=100$, the discount rate was set as $\gamma = 0.99$ and the learning rate $\alpha=0.0005$. We train over 400,000 time steps with target network updates frequency of $\tau=500$ steps. The size of the replay memory buffer is 50000 and the minibatch size is 32 (such that 1000 time steps are sampled from the buffer at each train iteration). We use an $\epsilon$ greedy policy with annealing wherein $\epsilon$ decreases linearly from 1 to 0.02 in 200,000 steps. 

\begin{figure}
\centering
\includegraphics[width=0.49\linewidth]{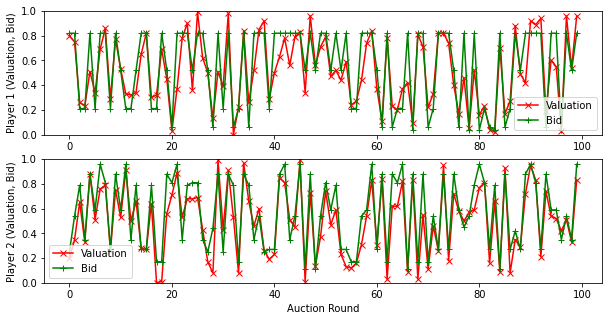}
\includegraphics[width=0.49\linewidth]{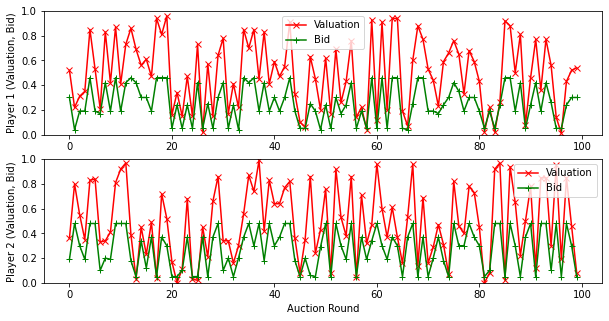}
\caption{The roll-out of the optimal bidding strategy of each bidder with deep state representations in second price auction (left) and first price auction (right).}
\label{fig:spa_past_rollout} 
\end{figure}

To capture the inter-dependence of actions chosen by the bidder, we model the observation for each agent as the current private valuation, the previous valuation, the previous bid and the auction outcome. Like before, we observe that the agents bid approximately with truthful bids for second price auctions and BNE for first price auctions in Figures~\ref{fig:spa_past_rollout}. 

\end{document}